# Causality and local determinism versus quantum nonlocality.

M Kupczynski

Département de l'Informatique, UQO, Case postale 1250, succursale Hull, Gatineau. QC, Canada J8X 3X 7

marian.kupczynski@uqo.ca

**Abstract**. The entanglement and the violation of Bell and CHSH inequalities in spin polarization correlation experiments (SPCE) is considered to be one of the biggest mysteries of Nature and is called *quantum nonlocality*. In this paper we show once again that this conclusion is based on imprecise terminology and on the lack of understanding of probabilistic models used in various proofs of Bell and CHSH theorems. These models are inconsistent with experimental protocols used in SPCE. This is the only reason why Bell and CHSH inequalities are violated. A probabilistic non-signalling description of SPCE, consistent with quantum predictions, is possible and it depends explicitly on the context of each experiment. It is also deterministic in the sense that the outcome is determined by supplementary local parameters describing both a physical signals and measuring instruments. The existence of such description gives additional arguments that quantum theory is emergent from some more detailed theory respecting causality and local determinism. If quantum theory is emergent then there exist perhaps some fine structures in time-series of experimental data which were not predicted by quantum theory. In this paper we explain how a systematic search for such fine structures can be done. If such reproducible fine structures were found it would show that quantum theory is not predictably complete what would be a major discovery.

## 1. Introduction.

Quantum theory (QT), giving accurate predictions for statistical distributions of outcomes obtained in physical experiments, is unable to tell which outcome and when will be observed. In spite of this it has been often claimed that QT provides the most complete description of individual physical systems and that *outcomes of quantum measurements are produced in irreducibly random* way.

According to statistical interpretation first advocated by Einstein [1] QT does not provide a complete description of individual physical systems and may be emergent from some more detailed theory of physical phenomena. If contexts of experiments are correctly taken into account this interpretation is free of paradoxes [2].

In particular the long range correlations predicted by QT for EPR type experiments and confirmed in spin polarization correlation experiments (SPCE) seem to *"cry for explanation"*.

Searching for an intuitive explanation of these correlations Bell analyzed local realistic and local stochastic hidden variable models. These models led to so called Bell and CHSH inequalities (CHSH) which were violated by some QT predictions [3, 4].

In *local realistic* models values of physical observables are predetermined by a source and recorded passively by a measuring apparatus. A notion of *local realism* is understood here in a narrow and specific sense. In stochastic models experimental outcomes are obtained locally in irreducibly random way.

A general belief was that no other local models were possible therefore the experimentally confirmed violation of CHSH [5,6] has been incorrectly interpreted as a mysterious non locality of Nature with correlations coming from out of space-time and paraphrased sometimes as: *"two perfectly random dice tossed in two far away locations produce perfectly correlated outcomes"*.

Of course a photon is not a die, correlations are not perfect and the confusion came from imprecise terminology and from the lack of understanding of probabilistic models used in various proofs of CHSH. All these models used a unique probability space and a joint probability distribution to describe all the results of different experiments performed in incompatible experimental settings. Non probabilistic models used predetermination of the results and/or *counterfactual definiteness*.

Several authors pointed out that these assumptions were very restrictive and if not used CHSH simply could not be proven [7-34]. Local probabilistic models for SPCE consistent with QT were constructed [8-10, 19, 21, 25-27, 32-33]. Experiments in various domains of science were found with correlations violating CHSH [18]. Several experiments in quantum optics and in neutron interferometry could be simulated event by event in a local and causal way [35, 36]. The references, we are giving in this paper, are by no means exhaustive. Each cited paper contains many additional references.

These results seem to be unknown, not understood or ignored by part of physical community.

A possibility of rational explanation of strong correlation of the outcomes of SPCE gives arguments in favor that QT is an emergent theory. We will show that this idea has many other experimentally testable implications.

In particular if QT is emergent then all pure quantum ensembles are mixed statistical ensembles and its various sub-ensembles may have slightly different properties [19, 20, 23, 25, 26]. Such differences can be detected by using purity tests which we discussed extensively many years ago in a different context [37-39].

Moreover some reproducible fine structures may exist in time series of data which are destroyed when a standard statistical analysis is done. To find these fine structures one has to study experimental data using more sophisticated statistical tools [39-41]. The existence of such structures would give not only convincing arguments that QT is emergent but it would also show that QT is not predictably complete [25, 39].

This paper is organized as follows.

In section 2 we show that if we insist on *irreducible randomness* of local experiments the most general probabilistic model is equivalent to the model found in stochastic hidden variables approach. Therefore the violation of CHSH shows only that the assumption of *the irreducible randomness* in QT is incorrect giving additional arguments that QT is emergent from some microscopic theory respecting *causality* and *local determinism*.

In section 3 we construct a local contextual probabilistic model for which it is impossible to prove CHSH. This model is able to reproduce all quantum predictions for SPCE. We also show that contrary to the general belief QT does not predict strict anti-correlations of outcomes in SPCE. At the end of this section we talk about a wonderful migration of monarch butterflies which would be impossible without some kind of causality and determinism in Nature.

In section 4 we discuss other experimental implications of the assumption that QT is an emergent theory. In particular we show some statistical tools which can be used to detect fine structures in time series of data not predicted by QT.

The last section contains conclusions.

**2. Probabilities and irreducible randomness.**

A probability is a complex notion [14] therefore we have to tell how we define it.

Let us consider a simple experiment consisting on flipping a coin $c_1$ by a flipping device $d_1$. We are repeating this experiment several times and we notice that knowing all preceding outcomes we are unable to predict what will be the next outcome. Besides we notice that if we increase the number of repetitions and calculate the relative frequencies of appearances of heads (H) or tails (T) then these frequencies approach 0.5. We say that *outcomes are produced in a random way with probability 0.5* and our experiment is called a *random experiment*. If we label H as 1 and T as -1 we introduce a random variable $X_1$ obeying a Bernoulli probability distribution B(0.5) defined by probabilities: $P(X_1 = -1)= P(X_1=1)=0.5$.

We were dealing here with a fair coin and with a fair device. If we took a coin $c_2$ which was crooked or/and a not fair flipping device $d_2$ we could observe that the probability $p$ of observing H was not equal to 0.5. In this case we would introduce a random variable $X_2$ obeying a probability distribution B($p$) defined by the probabilities: $P(X_2= -1)= p$ and $P(X_2=1)=1-p$.

Therefore a probability is neither a property of a coin nor a property of a flipping device. A probability is only a property of a random experiment [21-23]. Therefore we are dealing always with conditional probabilities [17-18] depending on the experimental context and we will write it explicitly when it is necessary. Since QT predicts the probabilities of various random experiments it is a contextual theory.

We concluded that the outcomes of coin flipping experiments were produced in a random way however we know that a motion of a coin obeys the laws of classical physics so if we knew all important parameters having impact on its motion we could in principle predict with certainty each final outcome. Of course we don't know and we don't control these parameters but this type of randomness is often considered less perfect and is called a classical or *reducible randomness*.

The fathers of QT postulated without proof that the randomness observed in quantum measurements and phenomena is more fundamental than in classical physics being intrinsic and irreducible. We will show that the violation of CHSH gives indication that this is not a case.

If both quantum and classical randomness are reducible we can only say that given experimental outcomes are produced in a random way if the time series of these outcomes passes with success all tests of randomness invented by statisticians. We regularly produce such series of "random" outcomes using generators of pseudo-random numbers but a person knowing the algorithm knows exactly what will be the next number of a series. Similarly consecutive digits of the decimal development of π pass with success all tests of randomness.

Let us consider now two far away laboratories performing experiments $x$ and $y$ respectively on two physical signals $S_1$ and $S_2$ produced by some source $S$. Let us assume that outcomes of the experiments $(x, y)$ are $(a, b)$ where $a=\pm1$ and $b=\pm1$. The outcomes of these experiments gathered during some period of time form two samples of data $\{a_1,a_2,\ldots a_n \ldots\}$ and $\{b_1,b_2,..b_n \ldots\}$. In mathematical statistics these outcomes are observations of two time series of random variables $\{A_1, A_2 \ldots A_n \ldots\}$ and $\{B_1, B_2 \ldots B_n \ldots\}$ called sampling distributions.

If all $A_i$ are independent and identically distributed (i.i.d) as some random variable $A$ and all $B_i$ are i.i.d as some random variable $B$ then outcomes of the experiments $x$ and $y$ are completely described by conditional joint probability distributions $P(A=a, B=b \mid x, y, S_1, S_2)$. Since the signals $S_1$ and $S_2$ can be correlated at the source:

$$P(A=a, B=b \mid x, y, S_1, S_2) \neq P(A=a \mid x, S_1) P(B=b \mid y, S_2) \qquad (1)$$

However *if S is sending a pure statistical ensemble of signals* for example each beam is composed of some *identical physical systems and if the outcomes are obtained in irreducibly random way* then *the random variables A and B are independent* and

$$P(A=a, B=b \mid x, y, S_1, S_2) = P(A=a \mid x, S_1) P(B=b \mid y, S_2) \qquad (2)$$

From (2) we obtain immediately the factorization of expected values $E(AB) = E(A) E(B)$.

If we insist on *irreducible randomness* of the act of measurement we may only obtain non trivial correlations if we assume that $S$ is sending a mixed ensemble of signals in which each couple $(S_1(\lambda_1), S_2(\lambda_2))$ is included with a probability $P(\lambda)$ where $\lambda=(\lambda_1,\lambda_2)$ are labels of different couples. We are talking about signals because our reasoning holds even if a signal is not a beam of some discrete physical systems. In this case instead of (2) we obtain:

$$P(a,b|x,y) = \sum_{\lambda \in \Lambda} P(\lambda) P(a|x,\lambda_1) P(b|y,\lambda_2) \qquad (3)$$

and

$$E(AB) = E(AB|x,y) = \sum_{\lambda \in \Lambda} P(\lambda)(E(A|\lambda_1,x)E(B|\lambda_2,y) \qquad (4)$$

Using (4) and choosing in an appropriate way $P(\lambda)$ we may prove that *Cov* $(A, B) \neq 0$. A probabilistic model defined by (3) describes the experiment $(x, y)$ as a mixed ensemble of pairs of *independent* random experiments labelled by $\lambda$. Correlations which may be created in this way are quite limited. Besides to estimate the probabilities in (3) we should follow a completely different experimental protocol than the experimental protocol used in SPCE. A more detailed discussion of this point is given in [27, 28].

To make this point clearer let us consider a simple experiment consistent with (3). Charlie is sending to Alice and Bob particularly chosen pairs of dice and let $x$ and $y$ be simply *rolling a die* experiments. Than in order to estimate the probabilities of various outcomes using (3) the following protocol may be used. Alice and Bob receive a first pair of dice and they roll their die many times to estimate corresponding probabilities. They proceed in the same way with all *correlated* pairs sent by Charlie and at the end they average all the obtained estimates. It does not resemble the SPCE protocol.

If we perform four different experiments $(x, y)$, $(x', y)$, $(x, y')$ and $(x', y')$ using the same signals prepared by $S$ and if all $|E(A|\lambda_1)|\leq 1$ and $|E(B|\lambda_2)|\leq 1$ then using (4) we can easily prove CHSH inequalities:

$$|E(AB) - E(AB')| + |E(A'B) + E(A'B')| \leq 2 \qquad (5)$$

which are violated in SPCE and in many experiments from other domains of science [18].

Equation (4) was used *in stochastic hidden variable models* and called *locality condition*. This terminology is the source of unjustified speculations about the non-locality of Nature.

Equation (5) can also be proven in a different way using *local realistic hidden variable models*. In these models outcomes of all experiments are predetermined and there exist a unique probability space and a joint probability distribution for all possible experiments $(x, y)$ [11, 14, 22, 23, 29]. The outcomes obtained in experiments $x$ and $y$ are evaluations of two functions $A(\lambda_1) =\pm 1$ and $B(\lambda_2) =\pm 1$. Therefore following Bell [2] we obtain:

$$E(AB) = \sum_{\lambda \in \Lambda} P(\lambda) A(\lambda_1) B(\lambda_2) \qquad (6)$$

Using (6) we can first prove Bell inequalities and deduce (5) as a corollary [3-4]. A very simple proof is given for example in the appendix II of [9].

In experiments $x$ and $x'$ one is measuring incompatible physical observables therefore all experiments $(x, y)$, $(x', y)$, $(x, y')$ and $(x', y')$ require different experimental settings. It was already known to Boole and Kolmogorov that a unique probability space and a joint probability distribution able to describe such experiments do not need to exist [13-16, 22-26, 33-34].

Since (5) is violated a correct probabilistic description of experiments has to be contextual imitating the contextual description provided by QT. In SPCE outcomes of measurements of physical observables $A$ and $B$ depend on a triplet $(S, x, y)$ defining a particular experimental context. In QT this experimental context is also described by a triplet.

Physical systems prepared by $S$ are represented by a wave function $\psi \in H_1 \otimes H_2$ or by a density operator $\rho$. The local physical observables $A$ and $B$ are represented by hermitian operators $\hat{A} \otimes I$ and $I \otimes \hat{B}$. Using this notation conditional expectation values $E(AB|\rho) = Tr\rho\hat{A}\hat{B}$ are calculated. *There is no reason for the existence of a functional relation between expected values* obtained in *incompatible experimental settings*.

One obtains (5) only if a physical state $\rho$ is a convex sum of separable states: $\rho = \sum_{i=1}^{k} p_i \rho_i \otimes \tilde{\rho}_i$ where $0 < p_i < 1$ because in this case [28]:

$$E(AB|\rho) = \sum_{i=1}^{k} p_i E(A|\rho_i) E(A|\tilde{\rho}_i) \qquad (7)$$

Of course entangled spin singlet state in SPCE is not a convex sum of separable states therefore probabilistic models leading to (5) cannot be used.

In the next section we will present a contextual local probabilistic model able to reproduce all QT predictions for SPCE.

## 3. Memory and local determinism in Nature.

In order to preserve correlations between signals created at the source some memory of these correlations has to be preserved when the signals reach measuring instruments and when the measurements are completed. To achieve it a measurement process instead of being irreducibly random has to be strictly deterministic and has to depend both on a signal and on a state of the instrument in a moment of measurement.

An outcome is not predetermined at the source but it is only known when a measurement is completed. Therefore a measured value of the physical observable is not an attribute of individual physical systems but it is a contextual property created during the act of measurement [21, 26, 42].

As we told in preceding section for experiments requiring incompatible experimental settings a unique probability space and a joint probability distribution able to describe these experiments do not exist and each couple of experiments (x, y) may be only described by a specific probability space $\Lambda_{xy}$.

Such type of contextual probabilistic description [15-17, 21-22, 25, 26] reconciles in some sense Einstein and Bohr. Einstein [1] claimed that QT might not provide a complete description of individual physical systems. Bohr insisted that any description of quantum phenomena is contextual depending on whole experimental setting [42].

In contextual probabilistic model (3) is replaced by:

$$P(a,b|x,y) = \sum_{\lambda \in \Lambda_{xy}} P(\lambda) P(a|x, \lambda_1, \lambda_x) P(b|y, \lambda_2, \lambda_y) \tag{8}$$

where $P(\lambda) = P(\lambda_1, \lambda_2) P_x(\lambda_x) P_y(\lambda_y)$, $\lambda = (\lambda_1, \lambda_2, \lambda_x, \lambda_y)$ and $\Lambda_{xy}$ is different for each pair of experiments. The correlations between signals created by S are coded in $P(\lambda_1, \lambda_2)$. To preserve the partial memory of these correlations we have to assume a local determinism thus $P(a|x, \lambda_x)$ and $P(b|y, \lambda_y)$ are 0 or 1.

Probabilistic models defined by (8) are consistent with the experimental protocol of SPCE and they can reproduce any correlations predicted by QT and violating CHSH. *Local choices of experimental settings made by experimenters are completely independent* and the order in which they are made has no influence on the correlations observed.

One may also define a different contextual model violating CHSH by replacing in (6) $\Lambda$ by $\Lambda_{xy}$ and adding parameters describing instruments [27].

Summation in equation (8) or in modified equation (6) can be replaced by integration over variables $\lambda$ without changing the conclusions [26, 27, 30-31]. There exist various proofs of (5) not using explicitly (4) or (6). However in all these proofs the contextual character of physical observables is neglected and a *counterfactual reasoning* is used [9, 14, 21, 22, 24, 29].

Since models defined by (8) contain probability distributions of $P_x(\lambda_x)$ and $P_y(\lambda_y)$ therefore they do not predict perfect anti-correlations for the outcomes of SPCE. They do not differ in this respect from QT. According to statistical interpretation the spin singlet wave function commonly used to describe the preparation of physical systems in SPCE does not give any information about individual outcomes but only it is a mathematical tool which together with appropriate operators can be used to predict statistical regularities in experimental data.

In SPCE observables A and B are the spin projections on two directions characterized by angles $\theta_A$ and $\theta_B$ respectively. Calculation for a singlet state gives the expectation value $E(AB|\psi) = -\cos(\theta_A - \theta_B)$. However directions of spin polarization analyzers are never sharp and are only defined by some small intervals $I_A$ and $I_B$ containing angles close to $\theta_A$ and $\theta_B$ respectively.

Therefore even if detection efficiencies were perfect and if we dealt with a perfect singlet spin state the prediction for the measured expectation values is only:

$$E(AB|\psi) = -\iint_{I_A I_B} \cos(\theta_1 - \theta_2) \, d\rho_A(\theta_1) \, d\rho_B(\theta_2) \qquad (9)$$

From (9) it is clear that, contrary to a general belief, QT does not predict strict anti-correlation of clicks in experiments measuring *A* and *B* because in a case of continuous variables the probabilities it provides should be treated as probability densities and integrated respectively. It was pointed out many years ago [19, 21]. More detailed discussion of this problem may be found in [24-26].

The existence of probabilistic models (8) able to reproduce predictions of QT seems to indicate that QT is an emergent theory and that there is no irreducible randomness in Nature.

One also may find many arguments in favour of determinism in biology. As an example we describe below *a wonderful migration* of monarch butterflies.

The monarch butterflies make massive, over 3000 km long, southward and northward yearly migrations from Ontario province in Canada to the sanctuaries of the Mariposa Biosphere Reserve in Mexico. In difference with birds no single individual makes the entire round trip.

The length of these journeys exceeds *a normal lifespan of monarchs which is around six weeks*. A normal life cycle of monarchs has 4 stages. From eggs they hatch into caterpillars, and then wrap up in cocoons where by *metamorphosis* they become butterflies. They get out from the cocoons, mate, lay eggs and die. The last generation of the summer, *with life span exceeding 7 months*, flies to one of many overwintering sites.

The overwintering generation does not reproduce until it leaves for the northward migration sometime in February and March and arrives as far north as Texas. Only the second, third or fourth generations return to their northern locations in the United States and Canada in late spring.

A true mystery is *how monarchs can return to the same overwintering spots over a gap of several generations*. The only possibility is that the *migration patterns are deterministically coded in their genes*. But even if these migration patterns are inherited it is still amazing how a small butterfly navigates using the sun, sky and earth's magnetic field to arrive exactly where it should. More information can be found for example on Wikipedia.

There are plenty other phenomena in biology which also indicate that: "*Nature does not play dice*" and that *irreducible randomness* is a false concept.

### 4. Is quantum theory predictably complete?

If QT is an emergent theory this fact has many model independent implications which in principle can be discovered by a detailed study of time series of experimental data.

In particular any model of sub-phenomena describes a pure quantum state as a mixed statistical ensemble with respect to some additional parameters. Any sub-ensemble of a pure ensemble is indistinguishable from the initial ensemble. The sub-ensembles of a mixed ensemble can differ between them. Since we do not control the distribution of supplementary parameters during the experiment time series of data may differ from run to run of the same experiment [19, 20, 43].

Therefore in an experiment in which it is believed that some beams are prepared in a pure quantum state one can make systematic changes of intensity and of geometry of the beams trying to change the properties of initial statistical ensemble. If the initial ensemble was mixed there is a chance to discover it.

To detect statistically significant differences between different experimental runs one has to use non parametric statistical tests which we studied many years ago and called *purity tests* [37, 38]. Since not many physicists know these tests we explain below two important tests: *run* and *rank tests*.

A *run* is a sequence of like elements that are preceded and followed by different elements or no elements at all. For example, in a sequence 00101100011011 we have eight runs and in a sequence 11111100000111 we have three runs. If the appearance of 1 and 0 is purely random the number of runs can neither be too small nor too big. If a sample size is $n = n_1 + n_2$, where $n_i$ are the numbers of 0 and 1 respectively, the test statistics to test the randomness of the series is $R = \#$ of runs. The expected value $E(R)$ and its variance $Var(R) = \sigma_R^2$ are functions of $n_1$ and $n_2$ and the probability distribution is known. For large values of $n$ a statistics $Z = (R - E(R))/\sigma_R$ is normally distributed.

In *rank tests* we rank the observations in increasing order and then we use the ranks for actual observations. For example in Mann–Whitney $U$ test we verify a null hypothesis that there is no significant difference between two populations. We combine two random samples S (1) and S (2) from these populations having sizes $n_1$ and $n_2$ respectively and we rank all the observations. To all tied observations the same averaged rank is assigned. The Mann–Whitney $U$ statistic is a function of $n_1$, $n_2$ and $R_1$ which is the sum of ranks scored by the observations from the sample S(1). For large samples one can use the standard normal distribution and tables exist for smaller samples.

A detailed discussion of these and other purity tests may be found in [37-39]. Using purity tests we may detect that data obtained in our experiment are a sample from some mixed statistical ensemble.

In order to answer the question whether QT is predictably complete we must study more carefully time series of experimental data in order to find some fine stochastic structures if they exist [39, 40].

Let us go back to our discussion of two samples of experimental data which we made on page 4 of this paper and consider data coming only from one experiment $x$. The outcomes of this experiment gathered during some period of time form a sample $\{a_1, a_2 \ldots a_n \ldots\}$ where individual outcomes are obtained in some discrete moments of time. In mathematical statistics these outcomes are observations a time series of random variables $\{A_1, A_2 \ldots A_n \ldots\}$.

If all $A_t$ are i.i.d as some random variable $A$ then outcomes of the experiment $x$ are completely described by a conditional probability distribution $P(A = a | x, S_1)$. *This is what is believed to be true for quantum measurements.* Therefore histograms and empirical probability distributions are compared with the predictions of QT and a fine structure of time series of data is not studied carefully enough.

If $A_t$ are not i.i.d random variables then experimental data are only completely described be the properties of the whole time series $\{A_t\}$.

A study of a time series is a delicate and a difficult task [39-41]. To detect a fine structure in a time series one has to use various models and try to match available data.

In particular for stationary time series a large class of autoregressive integrated moving average models ARIMA was studied extensively [41]. We define below a simplest autoregressive model *AR (p)* using a standard notation.

A time series it is a family of random variables $\{Z_t\}$ where $t=0,1..$ for simplicity. Time series is stationary if $E(Z_t)=\mu$, $Var(Z_t)=\sigma^2$ and covariance function $\gamma(k)$ at lag $k$ does not depend on $t$ where $\gamma(k) = Cov(Z_t, Z_{t+k}) = E((Z_t -\mu)(Z_{t+k} -\mu))$.

A white noise it is a time series $\{a_t\}$ where $a_t$ are normal i.i.d random variables with zero mean The autocorrelation function $\rho(k)$ at lag $k$ (*ACF*) is defined as $\rho(k) = \gamma(k) / \gamma(0)$ and it is easy to see that $\rho(0)=1$, $\rho(k) = \rho(-k)$ and $|\rho(k)| \leq 1$.

Autoregressive time series *AR (p)* with $\mu=0$ satisfies the following equation [40]:

$$Z_t - \Phi_1 Z_{t-1} - \Phi_2 Z_{t-2} - ... - \Phi_p Z_{t-p} = a_t \qquad (10)$$

where $\Phi_i$ are real numbers.

Using (10) one can prove that *ACF* is a quickly decaying function when *k* increases. Another important function having well known properties is so called partial autocorrelation function (*PAC*) which measures the importance of the lag *k* in *AR (p)*. These two autocorrelation functions are used to find out whether a studied time series can be modeled as *ARIMA* process and in particular as *AR (p)* process.

We cannot assume that our time series of experimental outcomes can be described using *AR (p)* model. Therefore we have to analyze carefully our data in order to find out which model if any is a correct one. In particular we have to study:

- Descriptive statistics.
- Histogram and normal scores plot.
- Simple time series plot $(z_t, t)$.
- Lagged scatter plots $((z_t, z_{t+k})$.
- Empirical *ACF* and *PAC* plots.

As we mentioned above a descriptive statistics and a histogram average out any fine structure of a time series. To illustrate this point we simulated a sample of size 500 of *AR* (2):

$$Z_t - 0.25 Z_{t-1} - 0.5 Z_{t-2} = a_t \qquad (11)$$

where $a_t$ were normal i.i.d. with a unit variance
.

A standard descriptive analysis: summary, histogram, and normal scores showed that our simulated data could be viewed as a sample drawn from some normally distributed population.

Using the statistical package $S^+$ we found that empirical *ACF* was decaying and empirical *PAC* had a clear <<cut-off>> at lag 2 allowing us to conclude the sample was drawn from a stationary *AR*(2) time series. Moreover the estimated values of the coefficients in equation (10) were: 0.243 and 0.487 very close to the true values 0.25 and 0.5 respectively [40]. It shows a power of statistical tools.

**5. Conclusions**

We have shown that one can explain *entanglement* and long range correlations without invoking *nonlocality* and *quantum magic*.

The violation of CHSH inequalities seems to indicate that *irreducible randomness* is a false notion and that "*Nature does not play dice*" supporting views of Einstein,'t Hooft [45] and many participants of this conference.

If QT is emergent from some detailed theory of physical phenomena then we can probably find confirmation of this idea by analyzing more carefully time series of experimental data.

Careful testing of CHSH inequalities was important because it allowed to check QT predictions and to eliminate probabilistic models which were unable to reproduce these predictions.

Since using local and contextual probabilistic models one cannot prove CHSH inequalities therefore further costly testing of these inequalities seems unnecessary. It would be better instead to search for reproducible fine structures in time series of data and answer the question whether QT is predictably complete [39, 40].

The contextual statistical interpretation of QT not only avoids various paradoxes but it allows also a consistent description of quantum measurement process [45].

According to this interpretation a wave function is not an attribute of a single physical state and it cannot be changed instantaneously. The reduced wave packet describes only the preparation of the sub-ensemble of the initial ensemble of physical systems for subsequent measurements.

Therefore there are no strict anti-correlations in SPCE and a registration of a particular click on Alice's detector does not give any *deterministic* prediction what happens or what will happen at Bob's detector. These facts are often misunderstood by members of quantum information community.

QT is extremely successful theory and perhaps it is the best description of physical phenomena we were able to find. Therefore even if QT is an emergent theory it does not mean that we will succeed to find more detailed consistent description of all observed physical phenomena.

However a possibility of explaining or simulating several quantum phenomena without referring to *quantum magic* is already an important step forward.

In spite of successes of QT there are still many worrying facts. In quantum field theory we introduce infinities by considering point-like charges and masses and we have to remove these infinities by using a complicated process of renormalization.

If one takes seriously the idea that hadrons are extended particles one may define a unitary S-matrix such that one cannot prove the *optical theorem* [46]. If the violation of the optical theorem was confirmed in LHC experiments it would require a reformulation of the standard model (SM).

Testing SM is a difficult task requiring many free parameters, various phenomenological inputs and Monte Carlo simulation of events. Even in quantum mechanics we have to use approximations and semi-empirical models containing free parameters in order to explain experimental data. It is worrying because in some sense QT becomes un-falsifiable [25, 43].

Let us conclude with two citations. One due to Bell [2]:"*What is proved by impossibility proofs is lack of imagination*" and one due to Bohr [42]: "*knowledge presents itself within a conceptual framework adapted to previous experience and …any such frame may prove too narrow to comprehend new experiences*".

Acknowledgments

I would like to thank Gerhard Grössing, Andrei Khrennikov, Theo Nieuwenhuizen, Kristel Michielsen and Hans de Raedt for interesting discussions.